\documentclass[a4paper,11pt]{article}
\pdfoutput=1 

\usepackage{jcappub} 

\usepackage[T1]{fontenc} 
\usepackage{caption}

\newcommand{\beq}{\begin{equation}}
\newcommand{\eeq}{\end{equation}}
\newcommand{\bea}{\begin{eqnarray}}
\newcommand{\eea}{\end{eqnarray}}
\newcommand{\beqn}{\begin{equation*}}
\newcommand{\eeqn}{\end{equation*}}
\newcommand{\bean}{\begin{eqnarray*}}
\newcommand{\eean}{\end{eqnarray*}}

\newcommand*{\cref}[1]{Chapter~\ref{#1}}

\DeclareMathOperator{\arcsech}{arcsech}

\title{\boldmath Model independent bounds for the number of $e$-folds during the evolution of the universe}

\author[a]{Gabriel Germ\'an}
\affiliation[a]{Instituto de Ciencias F\'{\i}sicas, Universidad Nacional Aut\'onoma de M\'exico, Av. Universidad s/n, Cuernavaca, Morelos 62210, Mexico}
\emailAdd{gabriel@icf.unam.mx}
\author[a]{R. Gonzalez Quaglia}
\emailAdd{rodrigo@icf.unam.mx}
\author[a,b]{A. M. Moran Colorado}
\affiliation[b]{Centro de Investigaci\'on en Ciencias, Universidad Aut\'onoma del Estado de Morelos, Avenida Universidad 1001, 
Cuernavaca, Morelos 62209, Mexico}
\emailAdd{abril.moranclo@uaem.edu.mx}

\abstract{We present a simple procedure to obtain universal bounds for quantities of cosmological interest, such as the number of $e$-folds during inflation, reheating, and radiation, as well as the reheating temperature. The main assumption is to represent each of the various epochs of evolution of the universe as being due to a single substance changing instantaneously into the next, describing a new era of evolution of the universe. This assumption, commonly used to obtain solutions of the Friedmann equations for simple cosmological models, is implemented here to find model-independent bounds on cosmological quantities of interest. In particular, we find that the bound $N_k\approx 56$ for $-\frac{1}{3} < \omega_{re} < \frac{1}{3}$ is very robust as an upper bound on the number of $e$-folds during inflation and also as a lower bound when $\omega_{re} > \frac{1}{3}$, where $\omega_{re}$ is the effective equation of state parameter during reheating. These are model-independent results that any single-field model of inflation should satisfy. As an example we illustrate with the basic $\alpha$ attractor model the usual model dependent approach, and the one presented here, and show how they complement each other.}

\begin{document}
\maketitle
\flushbottom

\section {\bf Introduction}\label{Intro}
Almost any textbook or review article on cosmology contains solutions of simple spatially flat cosmological models. These are solutions of the Friedmann equation that contain a single substance as the only component of the universe for a given epoch. Thus, for example, we have solutions with only radiation, only matter, etc. These simple toy models give a first understanding of the evolution of the universe with not too far solutions to precise solutions of more realistic models containing various substances. A similar strategy to the one described before can be followed to obtain approximate universal bounds for various quantities of cosmological interest, such as the number of $e$-folds during inflation, reheating, and radiation, as well as bounds for the reheating temperature (for reviews on inflation see e.g., \cite{Linde:1984ir}-\cite{Martin:2013tda} and  \cite{Bassett:2005xm}-\cite{Amin:2014eta}, for reheating).
The basic assumption, then, is to study the different epochs of evolution of the universe as due exclusively to the presence of a single substance at each epoch in instantaneous transition to the next. This assumption is good enough and is widely used. We will use it now to obtain universal bounds for cosmological quantities of interest although we understand that the approximate nature of the assumption will lead to small errors, quantified in Section \ref{Acc}, but the results should guide us correctly. Of particular interest are the upper bounds for the number of $e$-folds of observable inflation whose errors do not exceed one $e$-fold. 

The organization of the article is as follows: in Section \ref{MDB} we briefly review the standard procedure used to study the evolution of the number of $e$-folds of reheating and the reheating temperature as functions of the scalar spectral index $n_s$. This approach gives accurate results but has the disadvantage of being model dependent. In Section \ref{MIB} we present our approach which is model independent and sufficiently accurate to draw interesting conclusions from it. In Section \ref{Ex} we compare the results obtained from each of the two approaches in the basic $\alpha$-attractor model of inflation showing how both procedures complement each other. In Section \ref{Acc} we make a model independent assessment of the accuracy of the model independent (MIB) approach. Finally we conclude in Section \ref{CON}.
\section {\bf Model dependent bounds }\label{MDB} 
The usual approach to impose constraints to inflationary models coming from reheating starts with work done many years ago  \cite{Liddle:1994dx}, \cite{Liddle:2003as}, \cite{Dodelson:2003vq}, and culminating in an expression for the number of $e$-folds during reheating  \cite{Martin:2013tda}, \cite{Dai:2014jja}, \cite{Munoz:2014eqa}  as follows (see also e.g., section 3 of \cite{German:2020iwg})\footnote{The case $w_{re} = \frac{1}{3}$ must be treated independently as discussed in subsection 2.1 of \cite{Cook:2015vqa}}
\beq
\label{nre0}
N_{re}= \frac{4}{1-3\, \omega_{re}}\left(-N_{k}-\frac{1}{3} \ln[\frac{11 g_{s,re}}{43}]-\frac{1}{4} \ln[\frac{30}{\pi^2 g_{re} } ] -\ln[\frac{\rho^{1/4}_e k}{H_k\, a_0 T_0} ]\right),
\eeq
where $\omega_{re}$ is the EoS at the end of reheating,  $N_{k} \equiv \ln\left(\frac{a_e}{a_k}\right)$ is the number of $e$-folds during inflation after CMB scales on the pivot scale of wavenumber $k\equiv a_k H_k=0.05$/Mpc first left the horizon, the number of degrees of freedom of species at the end of reheating is denoted by $g_{re}$ and by $g_{s,re}$ the entropy number of degrees of freedom after reheating. The energy density at the end of inflation is $\rho_{e}$, with $a_0$ and $T_0$ the scale factor and temperature today, respectively. The energy density above is a model dependent quantity and can be written as $\rho_e=\frac{3}{2}V_e=\frac{9}{2}\frac{V_e}{V_k}H_k^2 M_{pl}^{2}=\frac{9\pi^2 A_s}{4}\frac{V_e}{V_k}r M_{pl}^{4}$, where $V_e$ is the potential of the model at the end of inflation, $V_k$ and $H_k$ are the potential and Hubble function at the comoving Hubble scale wavenumber $k$. 

The number of $e$-folds during radiation domination is given by
\beq
\label{nrd0}
N_{rd}= -\frac{3(1+\omega_{re})}{4}N_{re}+\frac{1}{4} \ln[\frac{30}{g_{re} \pi^2}] +\frac{1}{3} \ln[\frac{11 g_{sre}}{43}]+\ln[\frac{a_{eq}\, \rho_e^{1/4}}{a_0\,T_0}],
\eeq
where $a_{eq}$ denotes the scale factor at radiation-matter equality. These expressions are obtained by combining two basic equations, one which constrains the amount of expansion
\beq
\label{total}
\ln \frac{k}{a_0H_0} =\ln \frac{a_k}{a_e}\frac{a_e}{a_r}\frac{a_r}{a_{eq}}\frac{a_{eq}H_k}{a_0H_0}=- N_k - N_{re} - N_{rd}+\ln \frac{a_{eq}H_k}{a_0H_0}, 
\eeq
where  the number of $e$-folds during reheating is $N_{re} \equiv \ln \left(\frac{a_r}{a_e}\right)$, and during radiation $N_{rd}\equiv\ln\left(\frac{a_{eq}}{a_r}\right)$, defined as commonly used.  A second equation, together with considerations about entropy conservation after reheating, follows the postinflationary evolution of the energy density and temperature and is given by
\beq
\label{nre1}
N_{re} = \frac{1}{3\left(1+\omega_{re}\right)}\ln \frac{\rho_e}{\rho_{re}}\,.
\eeq
Thus, one of these three quantities, $N_k$, remains unconstrained and it is given only by specifying a model of inflation. From \eqref{nre1} also follows an equation for the reheat temperature
\beq
\label{TRE}
T_{re}=\left( \frac{30\, \rho_e}{\pi^2 g_{re}} \right)^{1/4}\, e^{-\frac{3}{4}(1+\omega_{re})N_{re}},
\eeq
where $\rho_{re}=\left(\pi^2 g_{re}/30\right) T_{re}^4$. Following \cite{Dai:2014jja} (for a small sample of papers see e.g., \cite{Cook:2015vqa}-\cite{Deng:2022bwn}), the study of Eq.~\eqref{nre0} for $N_{re}$ has been usually done without specifying upper bounds and plots in the $n_s$\,--\,$N_{re}$ plane extend to arbitrary high number of $e$-folds, with few exceptions (see e.g., \cite{German:2020cbw}). We will see in the following section how one can find model independent bounds for $N_{k}$, $N_{re}$ and $N_{rd}$, in particular, which any model of single field inflation must satisfy. 
\section {\bf Model independent bounds }\label{MIB} 
\begin{figure*}[h!]
\begin{center}
\captionsetup{format=plain,justification=centerlast}
\includegraphics[trim = 0mm  0mm 1mm 1mm, clip, width=15.5cm, height=13.cm]{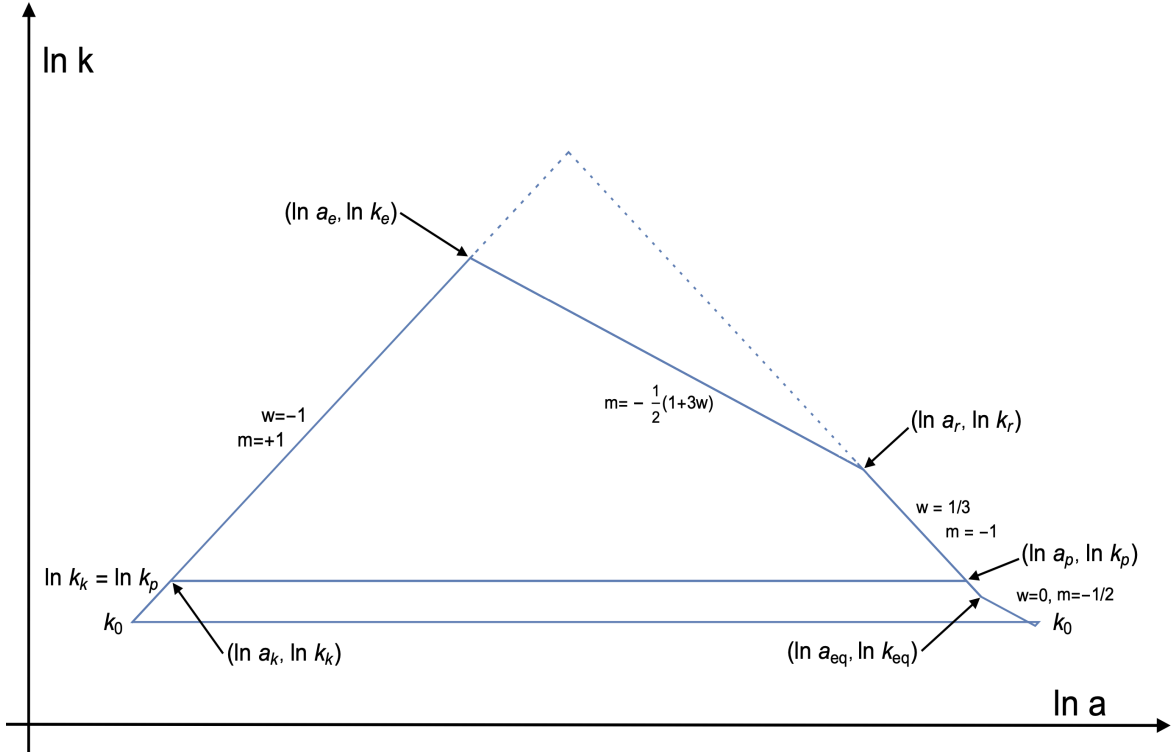}
\caption{\small Schematic diagram showing various epochs of the evolution of the universe with the logarithm of the comoving Hubble scale wavenumber mode  $\ln k$, where $k=a H$, as a function of the logarithm of the scale factor, $\ln a$. The comoving scale wavenumber $k$ exits the horizon during inflation (line of slope $m=+1$ in the l.h.s of the figure) and reenters  at the pivot scale $k_p=a_pH_p$ (line of slope $m=-1$ in the r.h.s of the figure) during radiation domination. The scale reenters at the pivot scale $k_p$ during the radiation era where the scale factor is $a_p < a_{eq}$, where $a_{eq}$ is the scale factor at radiation-matter equality. A line of slope $m=-\frac{1}{2}\left(1+3\omega_{re}\right)$ represents reheating, where $\omega_{re}$ is an effective equation of state parameter (EoS) assumed constant. Differences of their projections on the $\ln a$ axis determine the corresponding number of $e$-folds. The dotted lines can continue the inflation and radiation lines all the way to the vertex (by displacing the reheating line parallel to itself) where instantaneous reheating occurs with vanishing number of $e$-folds of reheating. Reheating lines with $\omega_{re}>1/3$ ($m < -1$) lie to the right of the radiation (dotted) line (not shown). }
\label{Dia1}
\end{center}
\end{figure*}

As a fast reference and for comparison with the model dependent (MDB) approach of last section we will call the one presented here as the MIB approach. We begin by establishing a very useful formula for the number of e-folds from the time a scale of wavenumber $k\equiv a_k H_k$ leave the horizon  to the time the same scale reenter the horizon at the pivot scale with wavenumber $k_p\equiv a_p H_p=0.05$/Mpc. This fixes the inflation line because $a_k= a_p e^{-N_{kp}}$ and $a_p$ can be calculated, for a given $k_p$, from the Friedmann equation. The equation for $N_{kp}$ is given by \cite{German:2020iwg} \footnote{We define the number of $e$-folds by $N_{ij}\equiv \ln\left(\frac{a_j}{a_i}\right)$, with the exception of the commonly used quantities like the number of $e$-folds during inflation $N_{k}$, reheating $N_{re}$, and radiation $N_{rd}$ defined before.}
\begin{equation}
N_{kp}\equiv \ln\left(\frac{a_p}{a_k}\right) = \ln[\frac{\, a_{p}\pi \sqrt{ A_s r}}{\sqrt{2}\,k_p}]\;,
\label{EQ}
\end{equation}
where $N_{kp}$ is the number of $e$-folds from the time scales leave the horizon when the scalar factor is $a_k$ to the  time these scales reenter the horizon at the pivot scale at $a_p$. Eq.~\eqref{EQ} above is simply obtained by multiplying $\frac{a_p}{a_k}$ above and below by $H_k$, realizing that $a_k H_k \equiv k=k_p$ and, from the amplitude of scalar perturbations at $k$, writing $H_k=\pi\sqrt{ r A_s/2}$. From \eqref{EQ} we have that $N_{kp}= N_{k}+N_{re}+N_{rp}$ with $N_{rp}\equiv \ln(\frac{a_p}{a_r}) =N_{rd} - \ln \frac{a_{eq}}{a_p}$, (see Fig.~\ref{Dia1}). 
In what follows we  determine independent formulas for $N_{re}$, $N_{rd}$ and from Eq.~\eqref{EQ}  we get an expression for $N_k$. We can visualize the problem by looking at Fig.~\ref{Dia1}. This diagram clearly indicates that the present study is reduced to the elementary problem of determining straight lines and their intersections.  Straight lines which follow from first solving the fluid equation for an assumed constant EoS $\omega$ 
obtaining $\rho\propto a^{-3(1+\omega)}$ and then making use of the Friedmann equation which can then be written as an equation for a straight line of the form $\ln k = m \ln a + b$ where $k=a H$ is the wavenumber mode associated with the comoving Hubble scale and $a$ is the scale factor of the universe.  The slope $m$ is directly related to the parameter $\omega$  as follows
\begin{equation}
m=-\frac{1}{2}\left(1+3\omega\right)\,.
\label{m}
\end{equation}
The radiation line is then given by $\ln k = - \ln a+b_r$, this line passes through the point $ \left(\ln a_p ,\ln k_p\right)$ (see Fig.~\ref{Dia1}) which, for a given $k_p$,  $a_p$ is obtained by solving Friedmann equation written in the form (in what follows we will set $a_0=1$)
\begin{equation}
 a_p = \frac{H_0\sqrt{\Omega_{rd,0}}}{k_p}\;,
\label{ap}
\end{equation}
where $\Omega_{rd,0}$ is the density parameter of radiation today. Usually we would write the Friedmann equation in the form $H\approx H_0\sqrt{\Omega_{md,0}/a^3+\Omega_{rd,0}/a^4}$, however, the nature of the approximation we are using (only requiring one type of substance for a given epoch) prevents this. Above, a curvature and a cosmological constant density parameters have been neglected.
Thus, the radiation line can be written as
\begin{equation}
\ln k = - \ln a+\ln\, H_0\sqrt{\Omega_{rd,0}}\;.
\label{ra}
\end{equation}
Proceeding in an analogous way the inflationary line, with the help of  \eqref{EQ}, can be written as $\ln k = \ln a+\ln \frac{k_p}{a_k }$ where $a_k = e^{-N_{kp}} a_p$, or
\begin{equation}
\ln k = \ln a+\ln \frac{\pi \sqrt{ A_s r}}{\sqrt{2}}\;,
\label{in}
\end{equation}
We join these two lines with a third one $\ln k = m \ln a+b_{re}$ with an arbitrary slope $m$ (arbitrary EoS). This line represents reheating leaving the inflationary line at $\left(\ln a_e ,\ln k_e\right)$ and reaching a point $\left(\ln a_r ,\ln k_r\right)$ in the radiation line. Imposing the condition on the reheating line that it intercepts the radiation line at the point $\left(\ln a_r ,\ln k_r\right)$ which signals the end of reheating or, equivalently, the beginning of the radiation epoch, gives $\ln k = m\ln a + \ln a_r^{-m} k_r$ where $a_r = e^{-N_{rp}} a_p$ and $k_r = \frac{H_0\sqrt{\Omega_{rd,0}}}{a_r}$. Thus,
\begin{equation}
\ln k = m\ln a + \ln \left[H_0\sqrt{\Omega_{rd,0}}\left(\frac{T_{re}}{\left(\frac{43}{11g_{s,re}}\right)^{1/3}T_0}\right)^{1+m}\right]. 
\label{re}
\end{equation}
and,
\begin{equation}
a_{r} =\left(\frac{43}{11g_{s,re}}\right)^{1/3}\, \frac{T_0}{T_{re}}\;.
\label{ar}
\end{equation}
A second condition on the reheating line comes from requiring that it also intercepts the inflationary line at  the end of inflation $\left(\ln a_e ,\ln k_e\right)$. We find 
\begin{equation}
a_{e} =\left(\frac{\sqrt{2}H_0\sqrt{\Omega_{rd,0}}}{\pi \sqrt{ A_s r}}\left(\frac{T_{re}}{\left(\frac{43}{11g_{s,re}}\right)^{1/3}T_0}\right)^{1+m}\right)^{\frac{1}{1-m}}\;.
\label{ae}
\end{equation}
These last two equations  allow the determination of $N_{re}$.
The number of $e$-folds during reheating from the end of inflation at $a_e$ to the beginning of radiation at $a_r$, is defined by $N_{re}\equiv \ln \frac{a_r}{a_e}$. Using Eq.~\eqref{m} to eliminate the slope $m$ in terms of the EoS $\omega_{re}$, we get 
\begin{equation}
N_{re} = \ln \left(\frac{\left(\frac{43}{11g_{s,re}}\right)^{2/3}\pi\sqrt{A_s r}\,T_0^2}{\sqrt{2}H_0\sqrt{\Omega_{rd,0}}\,T_{re}^2}\right)^{\frac{2}{3(1+\omega_{re})}}\;.
\label{ner}
\end{equation}
An upper bound for $N_{re}$ could be obtained, for each $\omega_{re}$, by considering the minimum possible value of $T_{re}$ simultaneously with the upper bound for $r$. 

\begin{figure*}[h!]
\captionsetup{format=plain,justification=centerlast}
\begin{center}
$\begin{array}{ccc}
\includegraphics[width=3.3in]{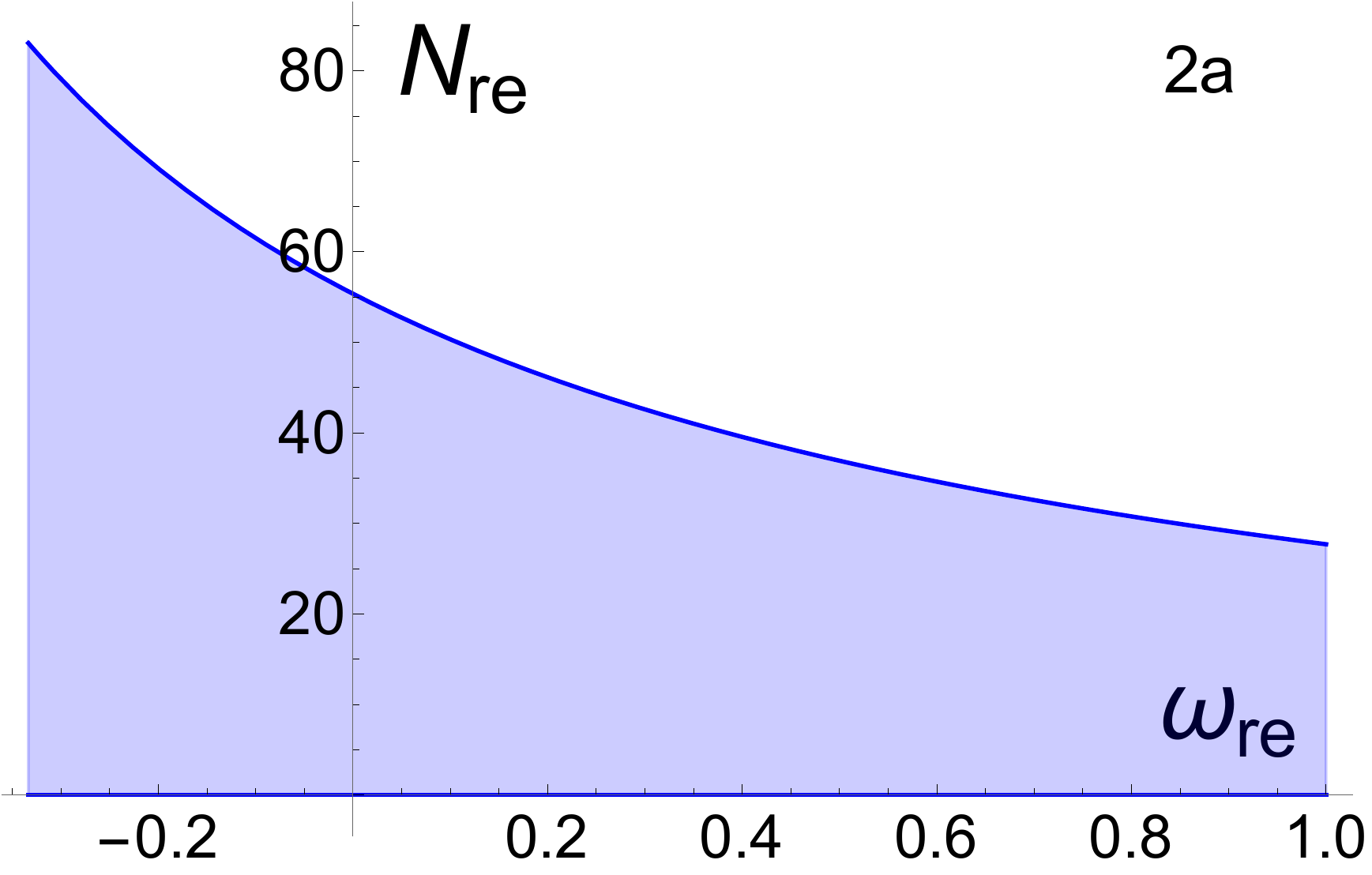}&
\includegraphics[width=3.3in]{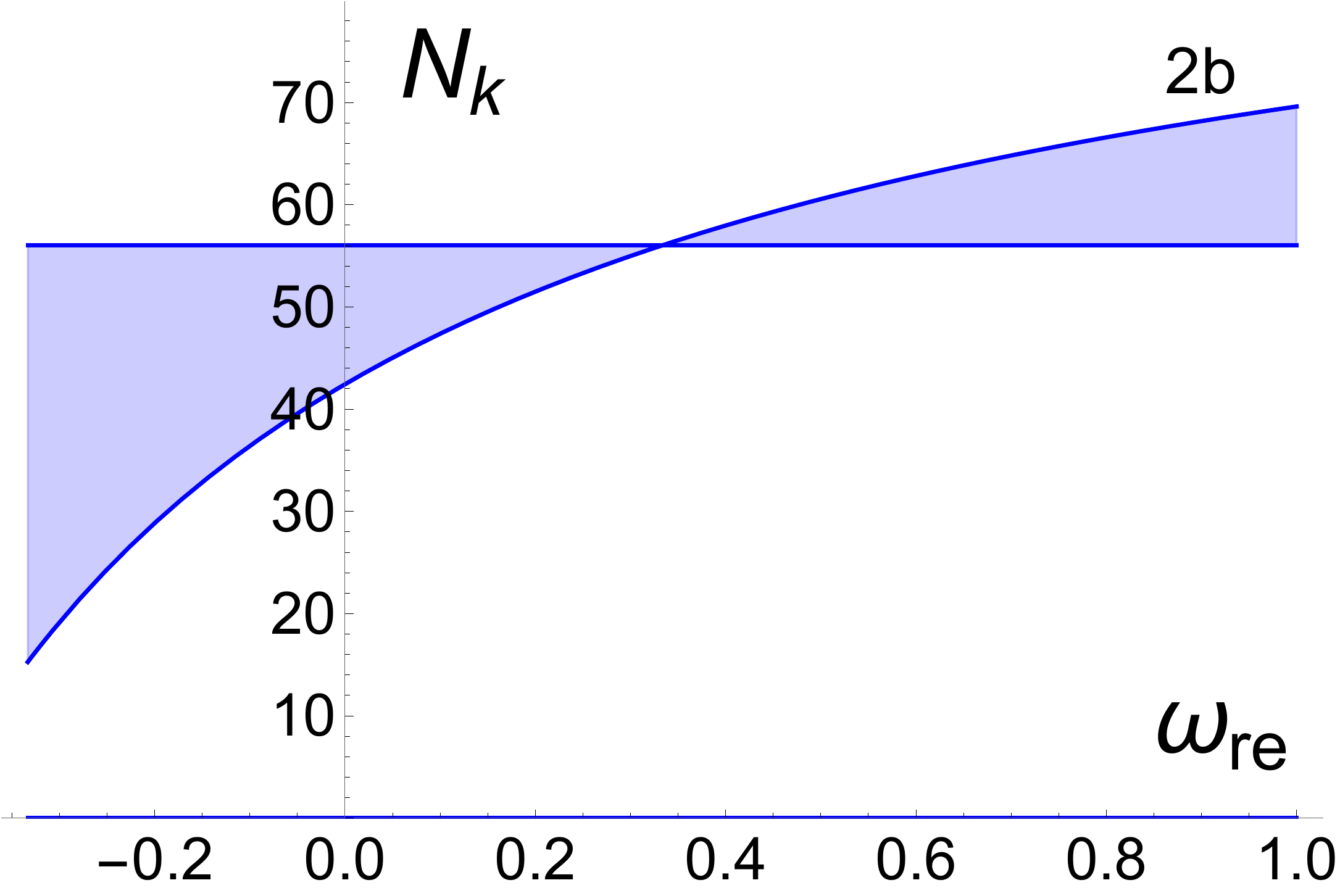}\\
\includegraphics[width=3.3in]{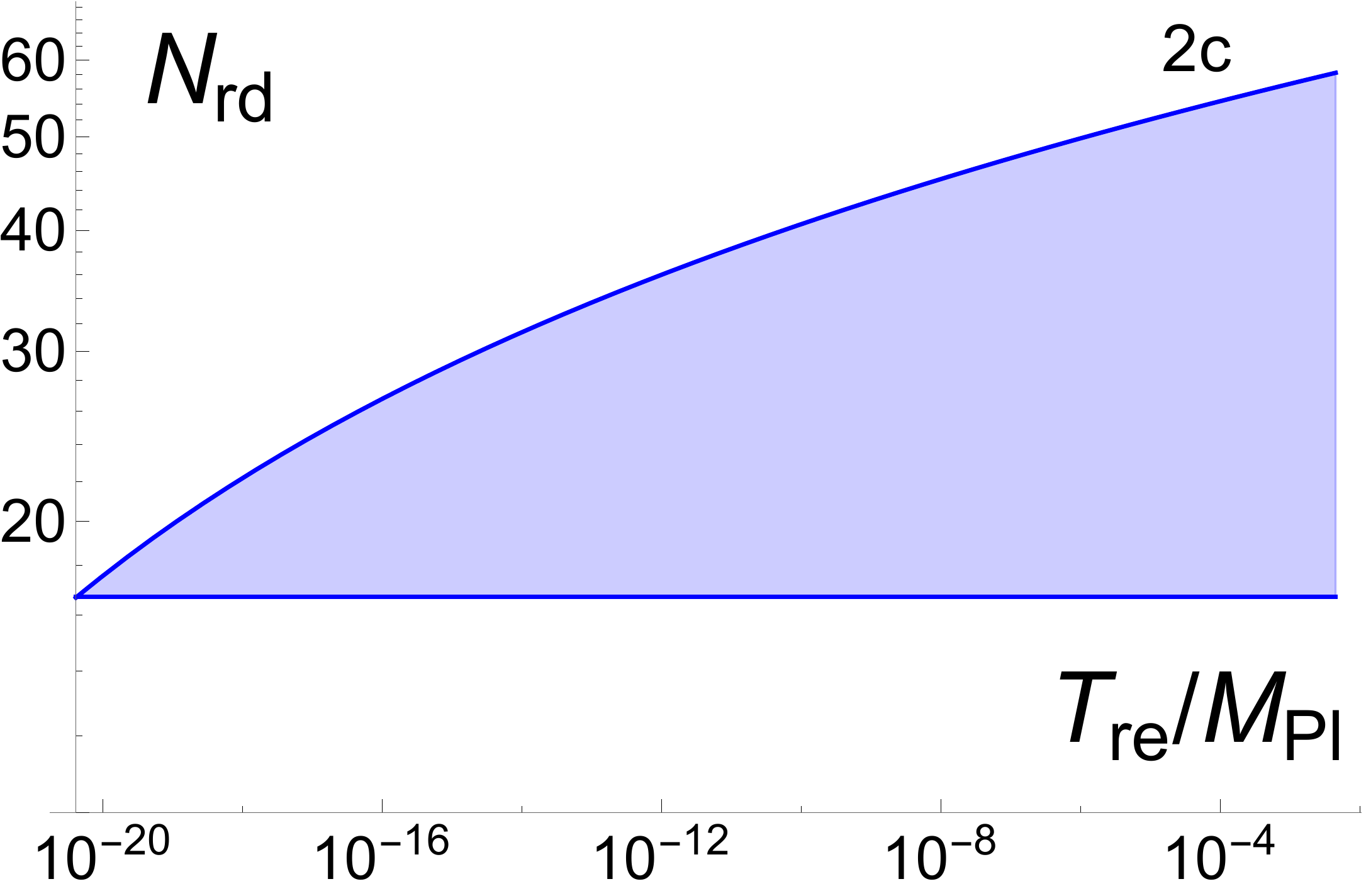}&
\includegraphics[width=3.3in]{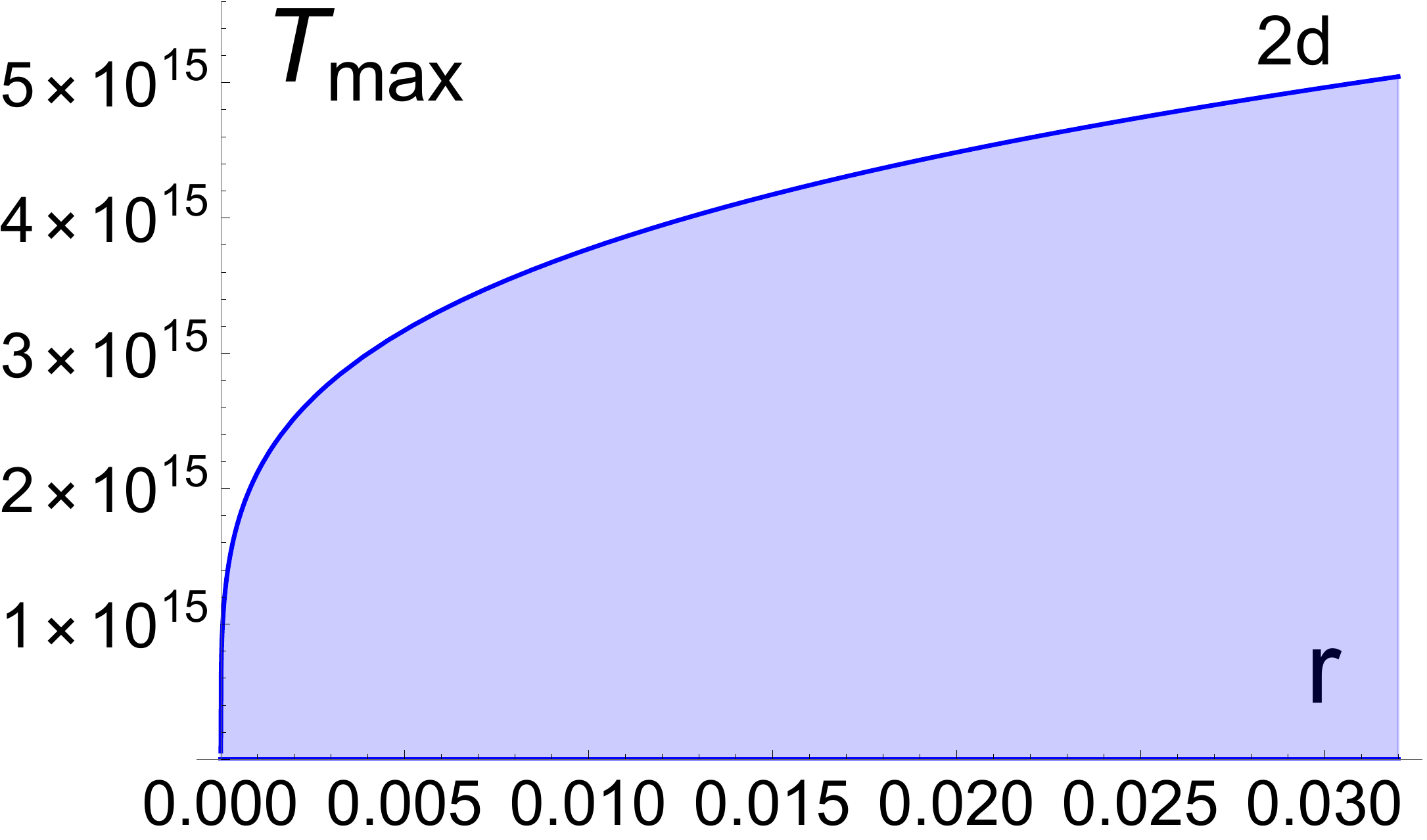} \\
\end{array}$
\caption{The figure 2a shows the number of $e$-folds of reheating Eq.~\eqref{ner} as a function of the  EoS $\omega_{re}$, from the maximum reheat temperature $T_{max}$ of Eq.~\eqref{tmax}, where $N_{re}=0$, to the maximum value of $N_{re}$ which occurs for the minimum reheat temperature $T_{min}$, here chosen as $10 MeV$ (see, however, \cite{Kawasaki:1999na}-\cite{Hasegawa:2019jsa}). There is no strong dependence on $r$ for values close to the upper bound $r_{0.05}=0.032$ \cite{Tristram:2021tvh}. Figure 2b is very interesting, it shows the number of $e$-folds during inflation Eq.~\eqref{nin} as a function of $\omega_{re}$ for $r=0.032$, between $T_{min}$ and $T_{max}$. For $T_{re}=T_{max}$, $N_{k}$ becomes  $\omega_{re}$-independent. We notice that for $-1/3<\omega_{re} < 1/3$ there is a universal upper bound $N_k<56$ which becomes a lower bound for $\omega_{re} > 1/3$. This behavior occurs for any value of $r$ because $N_{k}(T_{max}) =N_{k}(\omega_{re}=1/3)$ independently of $T_{re}$,  as illustrated in Fig.~\ref{Nkpanels}. The figure 2c shows the $\omega_{re}$-independent number of $e$-folds during radiation Eq.~\eqref{nrd} as a function of the normalized reheat temperature. Finally, we show the evolution with $r$ of the maximum reheat temperature $T_{max}$ given by Eq.~\eqref{tmax}, note that $T_{max}$ is $\omega_{re}$-independent. In the Table~\ref{bounds} we collect bounds for various quantities of interest for the special case of $\omega_{re}=0$ and also for a particular model of inflation.}
\label{Dia2}
\end{center}
\end{figure*}

Eliminating from Eq.~\eqref{nrd0} the model dependent part $\rho_e$ by using \eqref{TRE} we obtain the number of $e$-folds during radiation
\begin{equation}
N_{rd} =\ln\left(\frac{a_{eq}T_{re}}{(\frac{43}{11g_{s,re}})^{1/3}T_0}\right)\;.
\label{nrd}
\end{equation}
According with the Eq.\eqref{EQ}, we find that the number of $e$-folds during inflation is
\begin{equation}
N_{k} = \ln \left[\frac{\left(\frac{43}{11g_{s,re}}\right)^{1/3} \pi\sqrt{A_s r}\,T_0}{\sqrt{2}\,k_pT_{re}} \left(\frac{\sqrt{2}H_0\sqrt{\Omega_{rd,0}}\,T_{re}^2}{\left(\frac{43}{11g_{s,re}}\right)^{2/3}\pi\sqrt{A_s r}\,T_0^2}\right)^{\frac{2}{3(1+\omega_{re})}}\right]\;.
\label{nin}
\end{equation}
As can be seen from the previous equations, a minimum value of $T_{re}$ also implies a lower bound both for the number of $e$-folds during radiation and during the inflationary stage, while the value of $T_{re}$ that makes $N_{re}=0$ will give upper bounds for both radiation and inflation in a model independent way (see Fig.~\ref{Dia2}).
\begin{table*}[htbp!][h!]
\begin{center}
\captionsetup{format=plain,justification=centerlast}
\begin{tabular}{ccc}
Parameter & \quad Usually given as   &\,\, Dimensionless, used here\\ \hline 
& \quad   & \quad  \\[-2mm]
$H_0$ & \quad $100\,h\frac{km}{s}/Mpc$  & \quad $8.7581 \times 10^{-61}\,h$ \\[2mm]
$T_0$ & \quad $2.7255\, K$  & \quad $9.6423\times 10^{-32}$\\[2mm]
$A_s$ & \quad $2.1 \times 10^{-9}$  & \quad $2.1 \times 10^{-9}$\\[2mm]
$k_p$ & \quad $0.05/Mpc$  & \quad $1.3128\times 10^{-58}$\\[2mm]
$a_p$ & \quad $-$  & \quad $3.32\times 10^{-5}$ (MIB) or $3.65\times 10^{-5}$ (MDB)\\[2mm]
$\Omega_{md,0}$  & \quad $-$ & \quad $0.315$\\[2mm]
$\Omega_{rd,0}$   & \quad $2.47\times 10^{-5}h^{-2}$  & \quad $2.47\times 10^{-5}h^{-2}$\\[2mm]
$a_{eq}$  & \quad $2.94\times 10^{-4}$  & \quad $1.33\times 10^{-4}h^{-2}$\\[2mm]
$a_{0}$  & \quad $-$  & \quad $1$\\[2mm]
$h$  & \quad $-$  & \quad $0.674$\\[2mm]
$M_{Pl}$  & \quad $2.44\times 10^{18}$ GeV  & \quad $1$\\[2mm]
 \end{tabular}
 \caption{\label{parameters} For easy reference this table collects all the numerical values of parameters used in the paper. Dimensionless quantities have been obtained by working in Planck mass units, where $M_{pl}=2.44\times 10^{18} GeV$ is set to $M_{pl}=1,$ the pivot scale $k_p\equiv a_p H_p=0.05\frac{1}{Mpc}$ becomes a dimensionless number given by $k_p\approx 1.31\times 10^{-58}$. To calculate $a_p$ we have to specify $h$ for the Hubble parameter $H_0$ at the present time. We  take the value given  by Planck $h=0.674$ for definitiveness.  The solution of the Friedmann equation with just radiation in it gives $a_p$ in the MIB approach with value $a_p\approx 3.32\times 10^{-5}$ while in the MDB approach the Friedmann equation also contains a matter term (neglecting a cosmological constant and curvature terms) giving $a_p\approx 3.65\times 10^{-5}$.}
 \end{center}
\end{table*}
The maximum reheat temperature is reached when instantaneous reheating occurs ($N_{re} =0$) thus, from  \eqref{ner} follows that
\begin{equation}
T_{max} =\frac{\left(\frac{43}{11g_{s,re}}\right)^{1/3} \sqrt{\pi}  \left(A_s r\right)^{1/4}T_0 }{2^{1/4}\sqrt{H_0}\Omega_{rd,0}^{1/4}}\;,
\label{tmax}
\end{equation}
Equations \eqref{ner} - \eqref{tmax} are the main equations of the article, with them we can determine the number of $e$-folds during reheating, radiation and inflation as functions of the reheat temperature $T_{re}$ and the tensor-to-scalar ratio $r$ for a given EoS during reheating $\omega_{re}$ in a model independent way. Usually we do not know the reheat temperature but we have a lower bound required by nucleosynthesis. We can then have broad bounds (see Fig.~\ref{Dia2}) for these quantities which can then be used to constrain, or discard, models of inflation. In the Table~\ref{parameters} we collect numerical values of all quantities used in the calculations and plots. In the following section we study a particular example along these lines with the results compared with the MDB approach, given in the Table~\ref{bounds}, for the specific value $\omega_{re}=0$.

\begin{figure*}
\captionsetup{format=plain,justification=centerlast}
\begin{center}$
\begin{array}{ccc}
\includegraphics[width=3.3in]{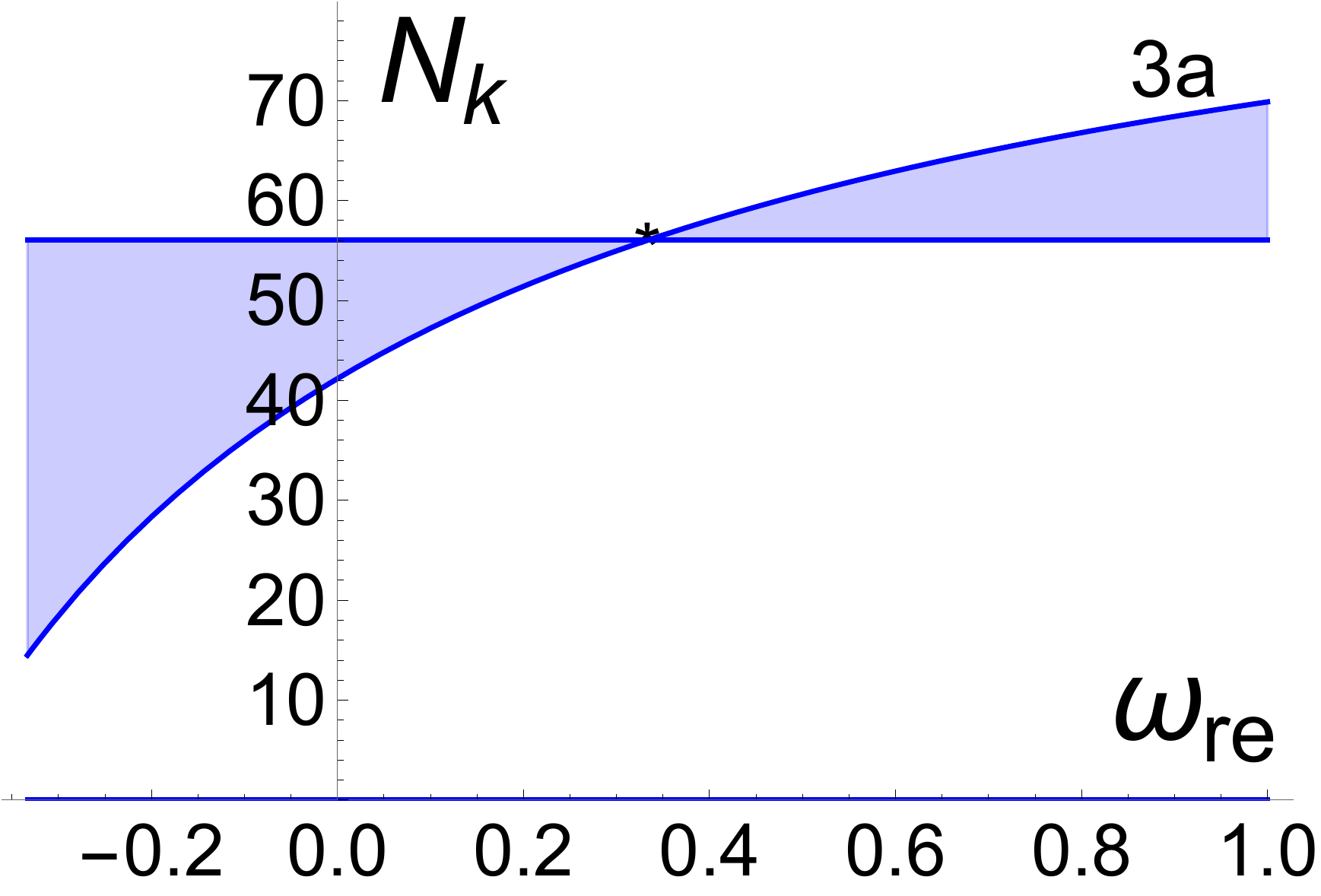}&
\includegraphics[width=3.3in]{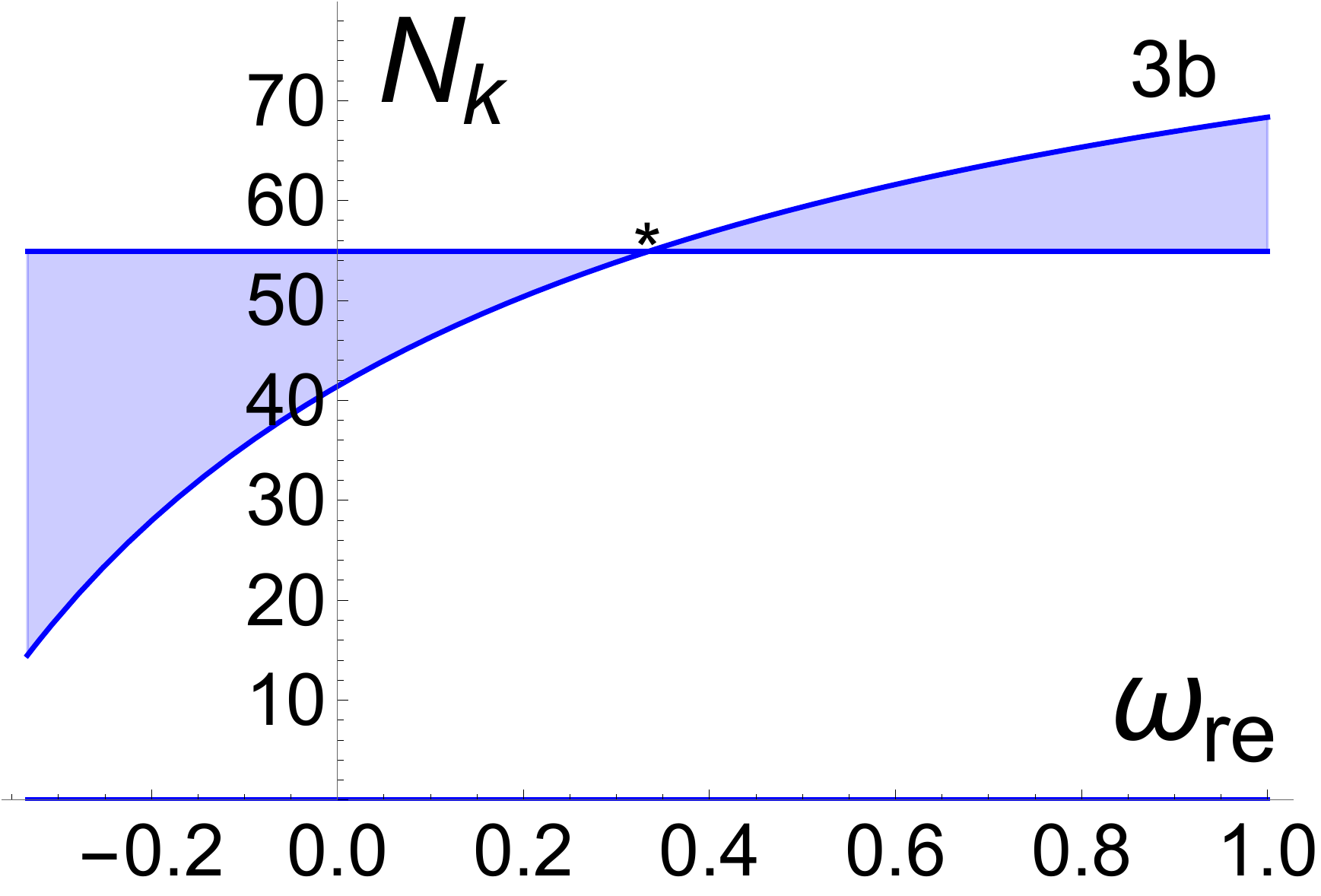}\\
\includegraphics[width=3.3in]{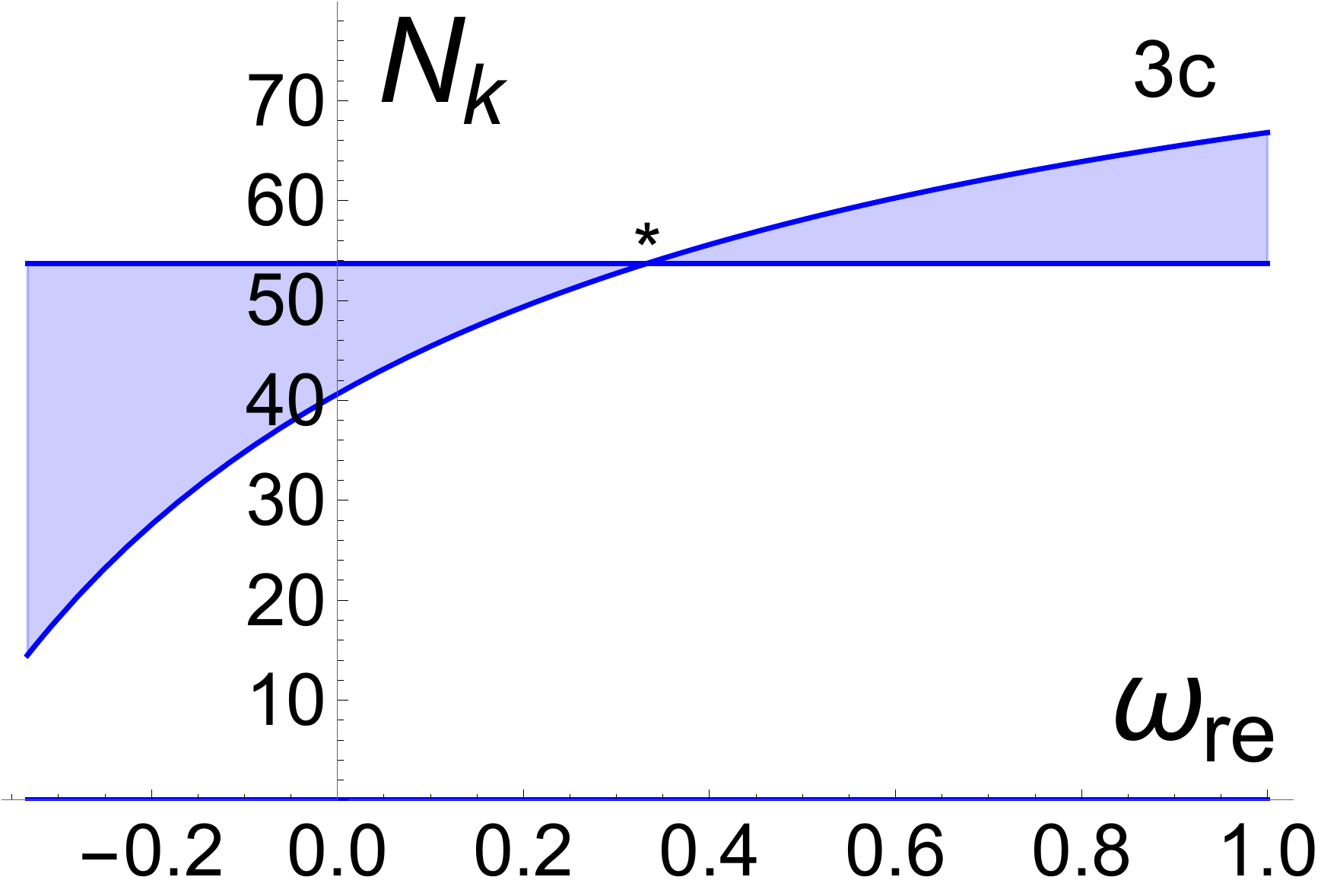}&
\includegraphics[width=3.3in]{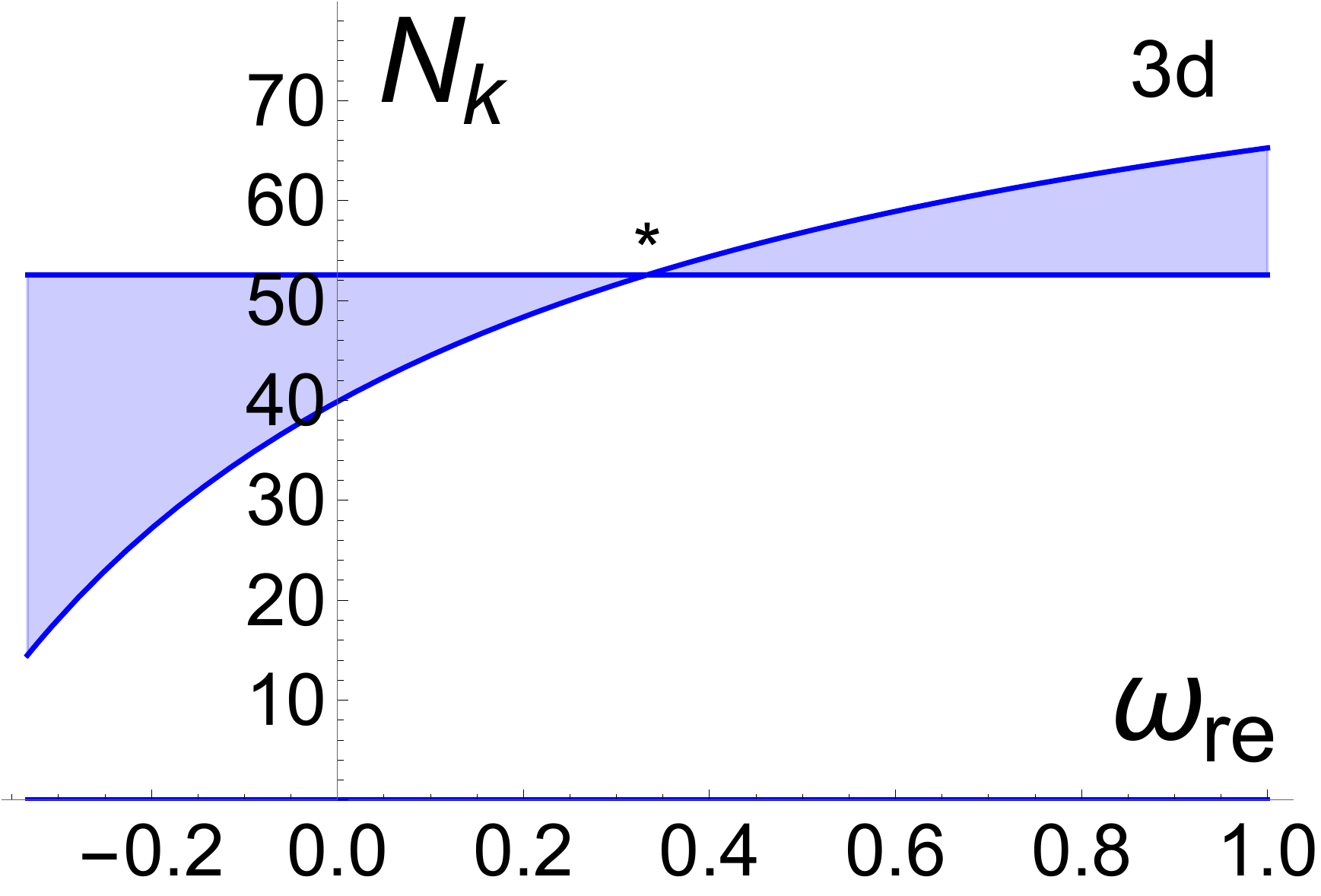} \\
\end{array}$
\caption{It can be shown that $N_{k}(T_{max}) =N_{k}(\omega_{re}=1/3)$ independently of $T_{re}$ thus, the lower bound and the upper bound curves always intersect at $\omega_{re}=1/3$ and various $N_k$ for any $r$ as shown in the different panels of the figure (see discussion after Eq.~\eqref{ninmax}). Panels 3a, 3b, 3c, and 3d are drawn for $r=3.2\times 10^{-n}$, $n$=2, 4, 6, and 8, respectively. The asterisk with coordinates $(1/3,56.0)$ is there for reference only to show the change of height of the curves at the intersection point.}
\label{Nkpanels}
\end{center}
\end{figure*}

Finally, let us note that for $T_{re} = T_{max}$ ($N_{re}=0$) the number of $e$-folds during inflation become $\omega_{re}$--independent
\begin{equation}
N_{k}(T_{max}) = \ln \left(\frac{\sqrt{\pi}A_s^{1/4}\sqrt{H_0\sqrt{\Omega_{rd,0}}}\,r^{1/4}}{2^{1/4}\,k_p} \right)\;,
\label{ninmax}
\end{equation}
and from Eq.~\eqref{ap} we see that $N_{k}(T_{max})=N_{kp}/2$ exactly, as could be expected from the symmetry of the large diagram (dotted lines) of Fig.~\ref{Dia1}. Also, from Eq.~\eqref{tmax} we see that, because $T_{max}$ cannot be lower that $T_{min}$ there must be an absolute lower bound for $r$ which is, however, extremely small $r_{min} \approx 2.3\times 10^{-74}$.
Finally, from Eq.~\eqref{nin} follows that $N_k\propto \ln T_{re}^{\frac{1-3\omega_{re}}{3(1+\omega_{re})}}$ thus,
we immediately see that for $\omega_{re}=1/3$, $N_k$ is independent of $T_{re}$ and even more, $N_{k}(\omega_{re}=1/3$) is exactly equal to $N_{k}(T_{max}) $. This means that for $\omega_{re}=1/3$ the curves $N_{k}(T_{max})$ and $N_{k}(\omega_{re}=1/3$) will intersect, as shown by Fig.~\ref{Nkpanels}, for any $r$.

\section {\bf Example: the basic $\alpha$ attractor model }\label{Ex}
\subsection {\bf The case $\omega_{re}=0$ }\label{Acc1}
In the interesting special case of $\omega_{re}=0$, the MIB approach immediately gives the broad (model independent) bounds
\begin{eqnarray}
\nonumber
&42.2<N_k<56.0\,,&\\
\nonumber
&55.4>N_{re}>0\,,&\\
\nonumber
&16.7<N_{rd}<58.2\,,&\\
&10MeV<T_{re}<5.0\times 10^{15}GeV,&
\label{bb} 
\end{eqnarray}
with a total expansion $N_{keq}=114.2$, from the time of horizon crossing to the time of the radiation-matter equality. The l.h.s. bounds are obtained by considering a minimal temperature during reheating $T_{min}=10MeV$  (see, however, \cite{Kawasaki:1999na}-\cite{Hasegawa:2019jsa}) and the current upper bound $r_{0.05}=0.032$ \cite{Tristram:2021tvh}, which gives a maximum number of $e$-folds during reheating. On the other hand, the r.h.s. bounds result from considering  $r_{0.05}=0.032$ and the maximum reheat temperature  $T_{max}$ (such that $N_{re}=0$). Note also that when calculating the bounds which follow from $T_{min} = 10 MeV$ we use $g_{s,re}=10.75=43/4$ because for that $T_{min}$ most $\pi^{\pm}$, $\pi^{0}$ and $\mu^-$ have annihilated leaving $e^{\pm}$, $\nu$, $\bar \nu$ and $\gamma$ behind. On the contrary the r.h.s bounds coming from $T_{max}$ are calculated using the value $g_{s,re}=106.75$ assuming that there is nothing new after the Standard Model of Particles. If we where considering some of its extensions e.g., the minimal supersymmetric standard model  then $g_{s,re}=915/4$ and so on. Because $g_{s,re}$ always occurs inside a $log$ it is of no great importance which value we use although, for definitiveness we take $g_{s,re}=106.75$ when $T_{re}\geq 200GeV$.
These bounds should be satisfied by any single field model of inflation where we can approximate the EoS during reheating by $\omega_{re}=0$ and with the caveat that there should be small errors due to the nature of the approximation used in the MIB approach. To have an idea of the magnitude of these errors let us consider in what follows a particular  example of current interest. The results are shown in the Table~{\ref{bounds} for both the MIB and the MDB approaches. 

Let us consider the basic $\alpha$-attractor potential \cite{Kallosh:2013hoa} given by 
\beq
\label{potanh}
V= V_0 \tanh^2\left(\frac{\phi}{\sqrt{6\alpha}M_{pl}}\right).
\eeq
Because we do not know the range of values the parameters $V_0$ and $\alpha$ can take, we eliminate them in terms of the observables $n_s$ and $r$. First we determine the value of $\phi$ at horizon crossing by solving for $\phi_k$ the equation for the amplitude of scalar perturbations $A_s(k) =\frac{1}{24\pi ^{2}\epsilon_k} \frac{V_k}{M_{pl}^{4}}$. Then we solve $16\epsilon_k=r$ for the parameter $\alpha$ and finally the expression for the scalar spectral index $n_{s} =1+2\eta_k -6\epsilon_k$ for the overall scale $V_0$. The results are \cite{German:2021rin}
\beq
\label{fik2}
\phi_k =\sqrt{\frac{8r}{\delta_{n_s}\left(4\delta_{n_s}-r\right)}}\arcsech\left(\frac{1}{2}\sqrt{\frac{r}{\delta_{n_s}}}\right)M_{pl}\,,
\eeq
\\
the parameter $\alpha$ is
\begin{equation}
\alpha= \frac{4r}{3\delta_{n_s}\left(4\delta_{n_s}-r\right)}\;,
\label{lambda2}
\end{equation}
and
\beq
\label{V02}
V_0 = \frac{6A_s \pi^2  r \delta_{n_s}}{4\delta_{n_s}-r}M_{pl}^4\,,
\eeq
where $\delta_{n_s}\equiv 1-n_s$. The end of inflation at $\phi_e$ is given by the condition $\epsilon=1$ where $\phi_e$ is 
\beq
\label{fie2}
\phi_e =\sqrt{
\frac{8r}{\delta_{n_s}\left(4\delta_{n_s}-r\right)}} \arcsech\left(
\frac{\sqrt{2}\,r^{1/4}\left(\sqrt{r+\delta_{n_s}\left(4\delta_{n_s}-r\right)}-\sqrt{r}\right)^{1/2}}{\sqrt{\delta_{n_s}\left(4\delta_{n_s}-r\right)}}\right)M_{pl}\,.
\eeq
The SR parameters appearing above are defined by
\begin{equation}
\epsilon \equiv \frac{M_{pl}^{2}}{2}\left( \frac{V^{\prime }}{V }\right) ^{2},\quad
\eta \equiv M_{pl}^{2}\frac{V^{\prime \prime }}{V}.
\label{SR}
\end{equation}
Also,  $M_{pl}\equiv 1/\sqrt{8\pi G}$ is the reduced Planck mass $M_{pl}=2.44\times 10^{18} \,\mathrm{GeV}$, primes on $V$ denote derivatives with respect to the inflaton field $\phi$.  
The number of $e$-folds during inflation is then given by
\beq
\label{Nk2}
N_{k} =\frac{8\delta_{n_s}-r-\sqrt{r^2+r\delta_{n_s}(4\delta_{n_s}-r)}}{\delta_{n_s}(4\delta_{n_s}-r)}.
\eeq
\begin{figure}[h!]
\begin{center}
\captionsetup{format=plain,justification=centerlast}
\includegraphics[trim = 0mm  0mm 1mm 1mm, clip, width=8.5cm, height=6.5cm]{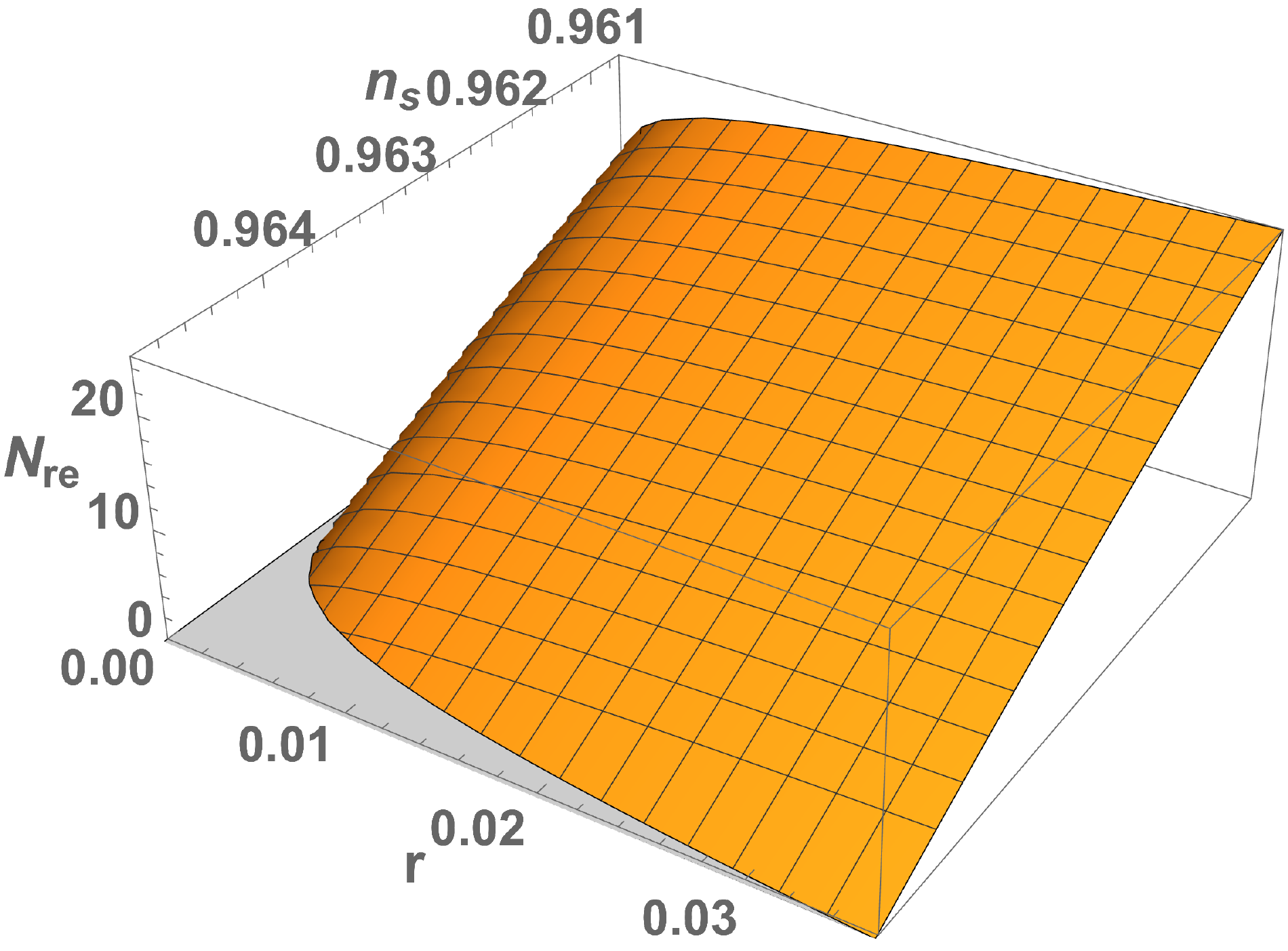}
\caption{Plot of the number of $e$-folds during reheating according to the MDB approach, Eq.~\eqref{nre0}, as a function of $n_s$ and $r$ for the model given by Eq.~\eqref{potanh}
and equation of state $\omega_{re}=0$. The original range for the tensor-to-scalar ratio is $r_{0.05} < 0.032$ \cite{Tristram:2021tvh}, and for the scalar spectral index $0.9607 < n_s < 0.9691$ \cite{Akrami:2018odb}. The condition $N_{re}\geq 0$ restricts the values of $n_s$ and $r$ further to the ranges $0.9607<n_s<0.9650$ and $8.9 \times 10^{-11} < r < 0.032$. From these bounds we obtain in turn new bounds for the number of $e$-folds during inflation $N_{k}$, radiation $N_{rd}$ and the reheat temperature $T_{re}$.}
\label{Nre}
\end{center}
\end{figure}
For the observables we consider first the bounds $r_{0.05}<0.032$ \cite{Tristram:2021tvh} and $0.9607<n_s<0.9691$ \cite{Akrami:2018odb}. We then impose the MIB constraint $N_k<56$ and as a consequence we find that $0.9607<n_s < 0.9646$. We also find for the model \eqref{potanh} a minimum value, $N_k=50.4$. With this value of $N_k$, $\omega_{re}=0$ and $r_{0.05}=0.032$ we find that $T_{re}=2.4\times 10^8$ GeV which becomes the new $T_{min}$ and from there we find the other bounds in the left hand side of $N_k$, $N_{re}$ and $N_{rd}$ given in the second column of the Table~\ref{bounds}. Although model dependent, these bounds have been obtained with the MIB approach just to compare with similar results obtained with the MDB procedure to have an idea of the discrepancies between the two approaches (third column of the Table~\ref{bounds}). Also from the MIB approach, we have seen that the universal bound $N_k<56.0$ constrains $n_s$ to the new range $0.9607<n_s<0.9646$ and from there we get constraints for the parameter $\alpha$ as shown also in the second column of the  Table~\ref{bounds}. Note that the {\it right hand side} bounds for $N_k$, $N_{re}$, $N_{rd}$ and $T_{re}$ are model independent because they follow from the original conditions $\omega_{re}=0$, a maximum temperature during reheating $T_{max}$ ($N_{re}=0$) and the upper bound $r_{0.05}=0.032$ which do not involve the $\alpha$-attractor model.

We now consider the MDB approach of section \ref{MDB}. Having determined $N_k$ by Eq.~\eqref{Nk2} it is more convenient to impose the bound $N_{re}=0$, corresponding to $T_{re}=T_{max}$, directly to Eq.~\eqref{nre0}, (see Fig.~\ref{Nre}). In this way we constrain the observables $n_s$ and $r$ to new ranges and with them constrain cosmological quantities of interest. The bounds so obtained are given in the third column of Table~\ref{bounds}. They are all model dependent. Note that in this more realistic approach we consider the full Friedmann equation $H=H_0\sqrt{\Omega_{rd,0}/a^4+\Omega_{md,0}/a^3+\Omega_{k,0}/a^2+\Omega_{\Lambda}}$ $\approx H_0\sqrt{\Omega_{rd,0}/a^4+\Omega_{md,0}/a^3}$ to get the relevant quantities. For example, the value of the scale factor at the pivot scale is now $a_p=3.65\times 10^{-5}$ compared to $a_p=3.32\times 10^{-5}$ where only one substance $H\approx H_0\sqrt{\Omega_{rd,0}/a^4}$ was considered for the MIB approach.
\begin{table*}[h!]
 \begin{center}
\captionsetup{format=plain,justification=centerlast}
{\begin{tabular}{cccc}
\small
Characteristic & MIB & MDB & $\delta$ \\ \hline\\[0.1mm]
$n_s$   &     ${\underline {0.9607}} < n_{s} < 0.9646 $ & ${\underline {0.9607}} < n_{s} < 0.9650$  & $\quad (0,+3.8\times 10^{-4}) $\\[2mm] 
$r$   &   ${\underline {0}}< r < {\underline {0.032}}$ & $8.89\times 10^{-11}< r <{\underline {0.032}} $ & $\quad (+8.89\times 10^{-11},0)$\\[2mm]
$ \alpha$   &   $ 0 < \alpha < 11.0$ & $1.9\times 10^{-8}  < \alpha < 11.3 $ & $ (+1.9\times 10^{-8},+0.3)$\\[2mm]
$N_{keq}$   &   ${\underline{114.2}}$ &   ${\underline{114.2}}$ & $0$\\[2mm]
$N_{k}$   &   $50.4 < N_{k} <{\underline {56.0}}$ &   $50.4 < N_{k} < 56.6$ & $(0,+0.6)$\\[2mm]
$N_{re}$  &    $22.5 > N_{re} > {\underline {0}}$ & $24.7 > N_{re} > 0$ & $(+2.2,0)$\\[2mm]
$N_{rd}$  &   $41.3 < N_{rd} < {\underline {58.2}}$ & $39.1 < N_{rd} < 57.6 $ & $(-2.2,-0.6)$\\[2mm]
$T_{re} (GeV)$  &  $\left(2.4\times 10^{8},\, {\underline {5.0\times 10^{15}}}\right)$ &  $\left(2.5\times 10^{7},\, 2.7\times 10^{15}\right)$  & $(9.6,1.8)$\\[2mm]
\end{tabular}}
\caption{\label{bounds} Comparison of MIB and MDB results for the $\alpha$-attractor model of Eq.~\eqref{potanh} for the special case of $\omega_{re}=0$. Underlined quantities are independent of the model of inflation used as example, all the other bounds depend on the particular model under consideration. Note that in the MIB approach there are model dependent quantities because the MIB formalism was applied to a particular model in order to compare results with the MDB approach. The last column shows the discrepancies between the two approaches. We see, in particular, that the discrepancy of the universal bound $N_k=56.0$ w.r.t. the 56.6 result of the MDB approach is very small. We expect this to be a general result with $N_k\approx 56$ a universal bound whenever $-1/3<\omega_{re} < 1/3$  becoming a lower bound for $\omega_{re} > 1/3$ as shown in the Fig.~\ref{Dia2}, panel 2b, and with an error no greater than one $e$-fold (see discussion in section \ref{Acc}). 
The MIB approach is useful to establish broad bounds obtained from $T_{min}$ and also from $T_{max}$ while in a particular model the MDB approach should be used for better results although, of course, they are not universal. The errors in the MIB approach come, of course, from assuming that only one substance is present in each epoch of evolution of the universe while in a more realistic approach various substances should be taken into account particularly near the transitions.}
\end{center}
\end{table*}
\subsection {\bf The case $\omega_{re}=\frac{1}{4}$ }\label{Acc2}
We now consider the case of $\omega_{re}=1/4$. As shown in \cite{Podolsky:2005bw}, the equation of state quickly approaches $w_{re} =1/4$ in the case of preheating (also tachyonic preheating, see  \cite{Dufaux:2006ee}) and stays there for a long period until full thermalization. Therefore,  $w_{re} = 1/4$ can be considered as a representative of non-perturbative reheating. The MIB approach immediately gives the broad (model independent) bounds
\begin{eqnarray}
\nonumber
&53.2<N_k<56.0\,,&\\
\nonumber
&44.3>N_{re}>0\,,&\\
\nonumber
&16.7<N_{rd}<58.2\,,&\\
&10MeV<T_{re}<5.0\times 10^{15}GeV,&
\label{bb} 
\end{eqnarray}
with a total expansion $N_{keq}=114.2$, from the time of horizon crossing to the time of the radiation-matter equality. Actually, the bounds for $N_k$ and $N_{re}$ for any $\omega_{re}$ in the range $-1/3<\omega_{re}<1$ can be extracted directly from Fig.~\ref{Dia2}. Following similar steps to subsection \ref{Acc1} we obtain the bounds presented in the Table~\ref{bounds2}.
\begin{table*}[h!]
 \begin{center}
\captionsetup{format=plain,justification=centerlast}
{\begin{tabular}{cccc}
\small
Characteristic & MIB & MDB & $\delta$ \\ \hline\\[0.1mm]
$n_s$   &     ${\underline {0.9607}} < n_{s} < 0.9646 $ & ${\underline {0.9607}} < n_{s} < 0.9650$  & $\quad (0,+3.8\times 10^{-4}) $\\[2mm] 
$r$   &   ${\underline {0}}< r < {\underline {0.032}}$ & $8.89\times 10^{-11}< r <{\underline {0.032}} $ & $\quad (+8.89\times 10^{-11},0)$\\[2mm]
$ \alpha$   &   $ 0 < \alpha < 11.0$ & $1.9\times 10^{-8}  < \alpha < 11.1 $ & $ (+1.9\times 10^{-8},+0.1)$\\[2mm]
$N_{keq}$   &   ${\underline{114.2}}$ &   ${\underline{114.2}}$ & $0$\\[2mm]
$N_{k}$   &   ${\underline{53.2}} < N_{k} <{\underline {56.0}}$ &   $54.1 < N_{k} < 56.6$ & $(0.9,+0.6)$\\[2mm]
$N_{re}$  &    ${\underline{44.3}}> N_{re} > {\underline {0}}$ & $43.4 > N_{re} > 0$ & $(-0.9,0)$\\[2mm]
$N_{rd}$  &   ${\underline{16.7}}< N_{rd} < {\underline {58.2}}$ & $16.7 < N_{rd} < 57.6 $ & $(0,-0.6)$\\[2mm]
$T_{re} (GeV)$  &  $\left({\underline{10^{-2}}},\, {\underline {5.0\times 10^{15}}}\right)$ &  $\left(10^{-2},\, 2.7\times 10^{15}\right)$  & $(0,1.8)$\\[2mm]
\end{tabular}}
\caption{\label{bounds2} Comparison of MIB and MDB results for the $\alpha$-attractor model of Eq.~\eqref{potanh} for the special case of $\omega_{re}=1/4$. Underlined quantities are independent of the model of inflation used as example, all the other bounds depend on the particular model under consideration. The last column shows the discrepancies between the two approaches.
We see that bounds for both $N_{re}$ and $N_{rd}$ are wider than in the $\omega_{re}=0$ case, while the bounds for $N_{k}$ are narrower but similar to the case $\omega_{re}=0$. We can easily understand this by using Eq.~(\ref{m}): for the case $\omega_{re}=0$ the slope of the reheating line is $m=-0.5$ which is actually shown by Fig.~\ref{Dia1}, while for the case $\omega_{re}=1/4$ the reheating line has slope $m=-0.875$. The straight line representing reheating in the latter case would appear more vertical, closer to the radiation line (dotted) allowing more $e$-folds during reheating and radiation while keeping the range of $N_{k}$ without much change. Note that, in terms of the diagram shown in Fig.~\ref{Dia1}, the number of $e$-folds of reheating approaches zero as the reheating line translates parallel to itself until which has a projection on the horizontal axis equal to zero when reaching the (dotted) vertex. Note also that the upper bounds are the same as for the case $\omega_{re}=0$ showing that these are $\omega_{re}$-independent. 
}
\end{center}
\end{table*}

\noindent In addition to models where $0\leq \omega_{re}  < 1/3$, known to the authors are reheating scenarios with $\omega_{re}  > 1/3$. These comprise the classical scenarios with flatter than quadratic potentials. If $V \propto \phi^p,$ then the coherent oscillation of the scalar gives $\omega_{re} = (p-2)/(p+2)$  \cite{Turner:1983he}, as an example see \cite{Garcia:2020wiy}. For $\omega_{re}=1$, models of the so-called stiff matter have been also studied e.g., \cite{Chavanis:2014lra}.
\section {\bf Accuracy of the results}\label{Acc}
Note from the third column of Table~\ref{bounds} that for the upper bound of $N_k$ there is a small difference with respect to the MIB result of less than one $e$-fold. We expect this to be the case for most models of inflation. This upper bound is obtained when $N_{re} = 0$ and from Eqs.~\eqref{nre0} and \eqref{nrd0}  we see that then
\beq
\label{Nk3}
N_{k} =-\frac{1}{3} \ln[\frac{11 g_{s,re}}{43}]-\frac{1}{4} \ln[\frac{30}{\pi^2 g_{re} } ] -\ln[\frac{\rho^{1/4}_e k}{H_k\, T_0} ]\,,
\eeq
and
\beq
\label{Nrp}
N_{rp} = \frac{1}{4} \ln[\frac{30}{g_{re} \pi^2}] +\frac{1}{3} \ln[\frac{11 g_{s,re}}{43}]+\ln[\frac{a_{p}\, \rho_e^{1/4}}{T_0}]\,,
\eeq
where, as before $N_{rp}=N_{rd}-\ln \frac{a_{eq}}{a_p}$ and that $\rho_e=\frac{3}{2}V_e =\frac{9}{2}\frac{V_e}{V_k}H_k^2M_{Pl}^2$.  Then
\beq
\label{dif}
N_k - N_{rp} =-\ln\left[\left(\frac{11 g_{s,re}}{43}\right)^{2/3}\left(\frac{135}{g_{re}\pi^2}\right)^{1/2}\left(\frac{a_pk_p}{T_0^2}\right)\right]+\frac{1}{2}\ln\left[\frac{V_k}{V_e}\right]\approx -0.5146+\frac{1}{2}\ln\left[\frac{V_k}{V_e}\right]\,.
\eeq
In the MIB approach the difference $N_k - N_{rp}$ is exactly zero when $N_{re} = 0$ and the bound $N_k = 56.0$ is a universal bound however, in the MDB procedure, we can see that for $V_k/V_e < 153$ half the difference $\frac{N_k - N_{rp} }{2}< 1$, i.e., $V_e$ has to be 153 times smaller than $V_k$ to obtain a difference between the number of $e$-folds of inflation and the universal upper bound ($N_k=56.0$) of one $e$-fold. It has to be half the difference because the extra amount of $e$-folds for $N_k$ has to be compensated by the corresponding decrease of $N_{rp}$ in such a way that $N_{kp}$ remains constant (see Table~\ref{bounds}). 
Note that the discussion above, although in the context of the MDB approach, is model independent because no potential has been specified. On the contrary, for the attractor model of the example $V_k/V_e<27.6$ and $\frac{N_k - N_{rp} }{2}=\frac{1.145}{2}\approx 0.6$ as shown in the Table~\ref{bounds}. The condition $V_k/V_e < 153$ should be met by most flat potentials. Thus, for potentials such that $V_k/V_e < 153$, we would immediately know that $N_k$ cannot be larger than the upper bound of 56+1 $e$-folds for $-\frac{1}{3} < \omega_{re} < \frac{1}{3}$ or smaller than the lower bound of 56+1 $e$-folds when $\frac{1}{3} < \omega_{re} $. This behaviour, in the purely MIB approach, is shown in the Fig.~\ref{Dia2}, panel 2b and also in Fig.~\ref{Nkpanels}, for various values of $r$.
\section {\bf Conclusions}\label{CON}
The most important conclusion of this work is that any single field model of inflation should satisfy the upper bound $N_k=56$ quite accurately, where $N_{k}$ is  the number of $e$-folds during inflation after CMB scales on the pivot scale of wavenumber $k=0.05$/Mpc first left the horizon, whenever the parameter of the equation of state during reheating $\omega_{re}$ is between the bounds $-1/3<\omega_{ re}<1/3$ and for $\omega_{re} >1/3$, $N_k= 56$ becomes a lower bound. The bound $N_k= 56$, either upper or lower, occurs for instantaneous reheating and $r_{0.05}=0.032$. This is a model-independent result and is illustrated in the Fig.~\ref{Nkpanels} for various values of $r$. Small errors associated with the nature of the assumptions are quantified in Section  \ref{Acc}.
Slightly less accurate results can be obtained for the bounds obtained away from instantaneous reheating but these are model dependent results and should be studied with the MDB approach of Section \ref{MDB} for better accuracy. Thus, both approaches complement each other rather nicely with the MIB approach providing the broad bounds corresponding to the extreme cases of the reheat temperature $T_{re}=T_{min}$ and $T_{re}=T_{max}$ and the MDB approach, once a model of inflation is proposed, giving more accurate results which are, however, not universal. 
In the Table~\ref{bounds} a comparison is made, on a particular model, between the model independent approach presented here and the usual model depend approach. Finally, we would like to point out that the above results are based on a very simple and commonly used idea to obtain solutions of the Friedmann equation, involving a single substance, for simple spatially flat cosmological models. This same idea reduces the Friedmann equation to that of a straight line and turns the problem discussed here into the elementary problem of the construction and intersection of straight lines. 
\section*{Acknowledgements}
RGQ would like to thank the National Council of Science and Technology (CONACyT) for its funding and support.

\end{document}